\documentclass[onecolumn,11pt]{article}
\usepackage{graphicx}
\usepackage{txfonts}
\usepackage{longtable}
\usepackage{lscape}

\begin{document}
  \begin{center}
  
   { \Large \bf
     Composition of the Interstellar Medium 
   }

   \bigskip
  
          by \\
   \bigskip
     { \large P. \ \ G n a c i \'n s k i \ \ 
                and \ \ 
              M. \ \ K r o g u l e c
        }
  
   \bigskip
  
      Institute of Theoretical Physics and Astrophysics, \\
              University of Gda\'nsk,
              ul. Wita Stwosza 57, 80-952 Gda\'nsk, Poland\\
             
              {e-mail: pg@iftia.univ.gda.pl, fizmkr@univ.gda.pl} \\
  
   \bigskip
   { \it Received ... }
   
   \date{}
   
   \bigskip
   
   ABSTRACT \\
   
   \end{center}
   
     We present an analysis of the Fe II, Ge II, Mg I, Mg II, S I, S II, Si II and Zn II interstellar lines for 63 stars. The column density of Mg II, S II, Si II and Zn II is well correlated with the distance. However, the column density -- distance relation should be used with care for the estimation of the distance to OB stars. For stars with large $f(H_2)$ this relation can lead to a large overestimation of the distance. Hydrogen -- normalised column densities of Mg II, Si II and Ge II (our largest samples of data) drop with the interstellar reddening E(B-V) as expected for elements that are incorporated into dust grains. The Ge II abundance (GeII/H) is lower in dense molecular clouds. The  abundances of all analysed elements are generally lower than their Solar System values.

   \bigskip
   

\section{Introduction}

  McKee \& Ostriker (1977) in their model have divided the interstellar medium (ISM) into three components: 
a low density hot ionised medium (HIM, n$\sim10^{-2.5}$ cm$^{-3}$, T$\sim10^{5.7}$ K), 
a dense cold neutral medium (CNM, n$\sim10^{1.6}$ cm$^{-3}$, T$\sim10^{1.9}$ K) and
a photoionized cloud corona (T$\sim8000$ K). The cloud corona itself was divided into two subregions:
warm ionized medium (WIM, fractional ionization x$_{WIM}\sim0.7$), and a
warm neutral medium (WNM, x$_{WNM}\sim0.1$). The filling factor of the cloud corona was estimated to be f$\sim0.2$, while the filling factor for the HIM was estimated to be f$\sim0.7 \div 0.8$. The cold, dense clouds fill only f$\sim0.02 \div 0.04$ of the interstellar space. Recently Cox (2005) wrote a review paper about the three--phase interstellar medium.

  Megier {\it et al.} (2005) pointed out, that the equivalent widths of the Ca II
interstellar absorption lines are very well correlated with the distance to the stars.
They proposed to use the Ca II H \& K
lines to measure the distances to the OB stars. They observed that the equivalent widths of the
K I and CH lines are less tightly related to the parallaxes (distances) of stars. They stated that the K I and CH that are present in cloud cores. This observation is in conformance with the McKee \& Ostriker model because the cold cloud cores have much lower filling factor that the cloud corona.
  
  An analysis of the interstellar Ti II, Ca II and Na I absorption lines was done by Welsh {\it et al.} (1997) and by Hunter {\it et al.} (2006). Column densities of the ionised Ti and Ca are well correlated, yielding a conclusion, that both elements reside in the same environment. Both Ti II and Ca II interstellar lines analysed by them are correlated with distance.
  
  We want to extend the results presented by Megier {\it et al.} (2005) and include to the
analysis elements observed in the ultraviolet spectral range. Because of the difficulties with  saturated lines we have used column densities instead of equivalent widths.
The Hubble Space Telescope (HST) spectra were used to calculate column densities of Fe II, Ge II, Mg I, Mg II, S I, S II, Si II, Zn II elements. Both GHRS (Goddard High Resolution Spectrograph) and STIS (Space Telescope Imaging Spectrograph) spectra were used.
  
In the next paragraph we present the method used to determine the column densities. The correlation between column densities and distances is analysed in \S3. The relationships between interstellar gas-phase abundances and interstellar reddening E(B-V) or f(H$_2$) are also presented in \S3. The conclusions are summarised in \S4.

\section{Column Densities}

 The spectra from the ultraviolet spectral range were downloaded from the HST Data Archive. The GHRS spectra taken in the FP-SPLIT mode were processed with IRAF tasks {\it poffsets} and {\it specalign} to achieve the final spectrum. When many spectra for one star were available the column density was calculated independently for each spectrum, and an average column density was determined.  The column densities were derived using the profile fitting technique. The absorption lines were fitted by Voigt profiles. The transitions for which the natural dumping constant ($\Gamma$) in not known (Mg II 1240 \AA\ doublet, Ge II 1237 \AA, Mg I 1829 \AA ) were fitted with a Gauss function. The cloud velocities (v), Doppler broadening parameters (b) and column densities (N) for multiple absorption components were simultaneously fitted to the observed spectrum. Both lines of magnesium doublet (at 2800 \AA\ or 1200 \AA) were also fitted simultaneously - v, b and N were common for both lines in the doublet. The wavelengths, oscillator strengths (f) and natural damping constants ($\Gamma$) were adopted from Morton (2003). 
 
  A convolution with a point spread function (PSF) was also performed. The PSF for the GHRS consists of two Gaussian components. The "core" Gaussian has a FWHM=1.05 diodes, while the "halo" component has FWHM=5.0 diodes (Spitzer \& Fitzpatrick, 1993). The relative contribution of the "core" and "halo" components into the PSF is wavelength dependent and was interpolated from the table given by Cardelli {\it et al.} (1990). The Gaussian PSF for the STIS spectrograph depends on wavelength, slit and the mode of observations. The tables with FWHM for the combination of mode and slit can be found in "STIS Instrument Handbook" (Kim Quijano {\it et al.}, 2003). The FWHM of the Gaussian PSF was wavelength-interpolated from these tables. The derived column densities are presented in Table \ref{CD}.
  
  The interstellar absorption lines of Fe II, Zn II and S II are often saturated. 
Therefore the column density of these elements was measured only in a few directions.
The results for these three elements are fragmentary, because in the directions
where the absorption lines were saturated, the column densities were not determined.

\section{Results}

\begin{figure*}[tbh]
   \centering
   \includegraphics[width=120mm]{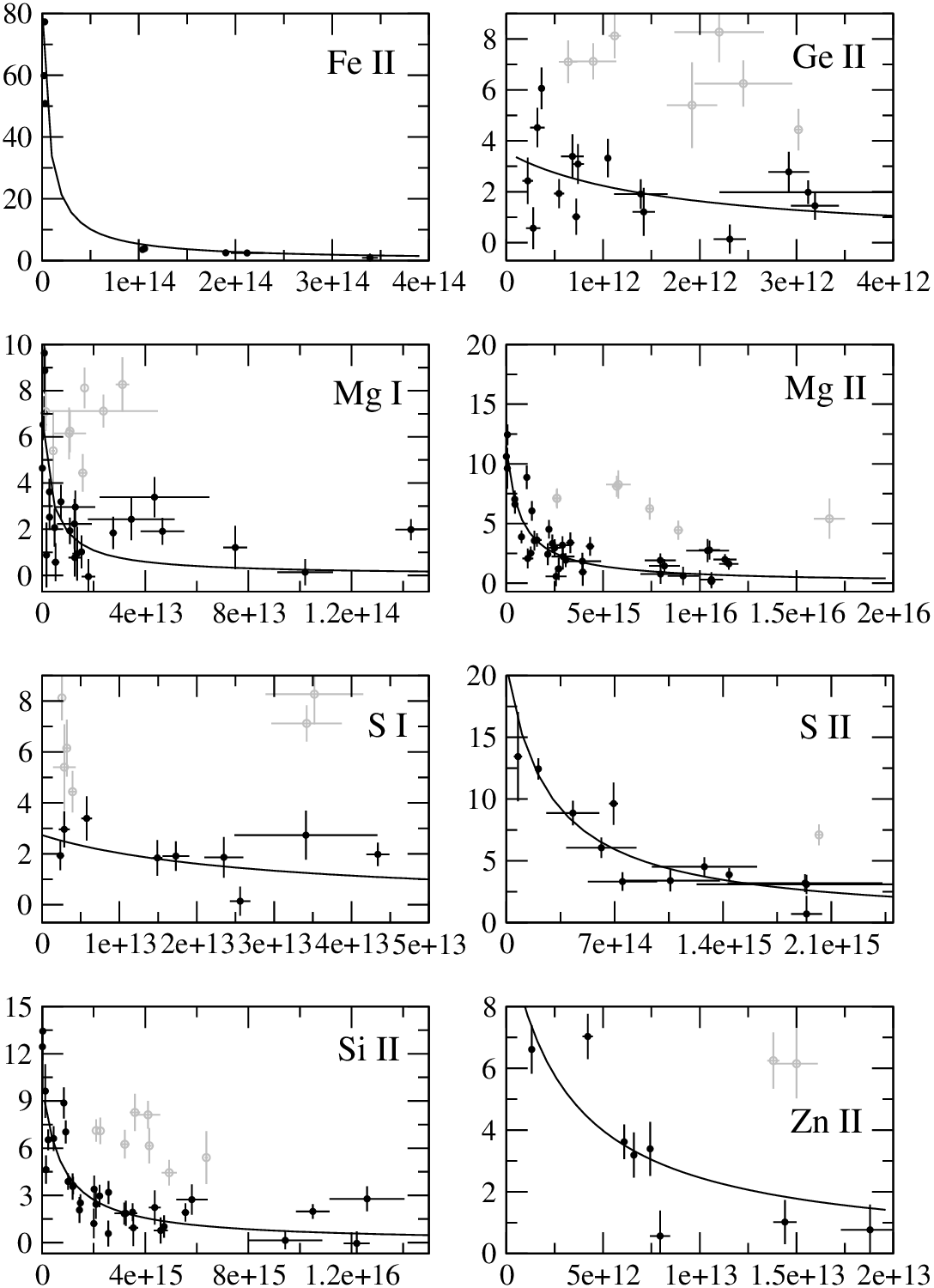}
      \caption{ Column densities [cm$^{-2}$] versus parallaxes [mas].
                The grey open circles represents stars from the Scorpius --  Ophiuchus
                region.
              }
      \label{paralax}
\end{figure*}

  The column densities of Fe II, Ge II, Mg I, Mg II, S I, S II, Si II, Zn II versus parallaxes are plotted  on Fig. \ref{paralax}. The parallaxes of stars used in this paper are taken from the Hipparcos Catalogue (1997).
  
  In order to check how the column density (N) changes with distance (d) a hyperbola was fitted (see Megier {\it et al.} (2005))
\begin{displaymath}
  \pi=\frac{1}{d}=\frac{1}{a \cdot N + b}
\end{displaymath}
in the column density - parallax ($\pi$) plane. We do not use the distance, like
Hunter {\it et al.} (2006) or Welsh {\it et al.} (1997), because only for the parallaxes the plus and minus errors 
are equal. Moreover, the parallaxes are sometimes negative. Removing such points from the sample leads to the underestimation of distance. The weight for each point in the fitting
procedure was equal to $1/\sigma_{\pi}^2$. 

The stars from the Scorpius-Ophiuchus region: HD 141637, HD 143018, HD 144217, HD 147165, HD 147933, HD 148479 belonging to the Upper Scorpius Association (de Zeeuw {\it et al.}, 1999), and HD 143275, HD 149757 were eliminated from the N-$\pi$ relation fitting. These are nearby stars lying behind a dense molecular cloud. Therefore these directions have very high column densities and deviate from the N-$\pi$ relation for other stars. The eliminated stars are shown with
open circles on Fig. \ref{paralax}. The seven stars from the local interstellar medium (LISM) with $\pi>20$ mas were also excluded from the fitting.
The results of the column density - parallax fitting are following:

\begin{math}
\begin{array}{lcl}
  d=(1.23 \pm 0.16 ) \cdot 10^{-13} \cdot N(MgII) + (89 \pm 8) & [pc] \\ 
  d=(1.72 \pm 0.25 ) \cdot 10^{-13} \cdot N(SII)  + (48 \pm 10) & [pc] \\ 
  d=(1.35 \pm 0.22 ) \cdot 10^{-13} \cdot N(SiII) + (97 \pm 11) & [pc] \\ 
  d=(3.1 \pm 1.2 ) \cdot 10^{-11} \cdot N(ZnII) + (96 \pm 38) & [pc] \\ 
\end{array}
\end{math} \\
where column densities (N) are in $cm^{-2}$, and distances are in parsecs.
There are no nearby stars with low column density for neutral elements (Mg I, S I) as well as for Ge II. Therefore the hyperbola fit gives very high $b$ term, that should be equal to the Local Bubble dimension. These fits are shown on Fig. \ref{paralax}, but these fits are probably meaningless.

  Megier {\it et al.} (2005) have proposed to use an analogous formula for Ca II to determine the distances to OB stars. In our sample of OB stars objects behind molecular clouds strongly deviate from the column density -- distance relation. For example the distance to HD 147933 from the Sco-Oph cloud calculated from the Mg II column density is 6 times grater (800 pc vs. 120 pc) than the distance determined from the Hipparcos parallax !
Such outliers have $f(H_2)>10^{-2}$, but only some stars with large $f(H_2)$ deviate from the N-$\pi$ relation.
  
\begin{figure*}[tbh]
   \centering
   \includegraphics[width=120mm]{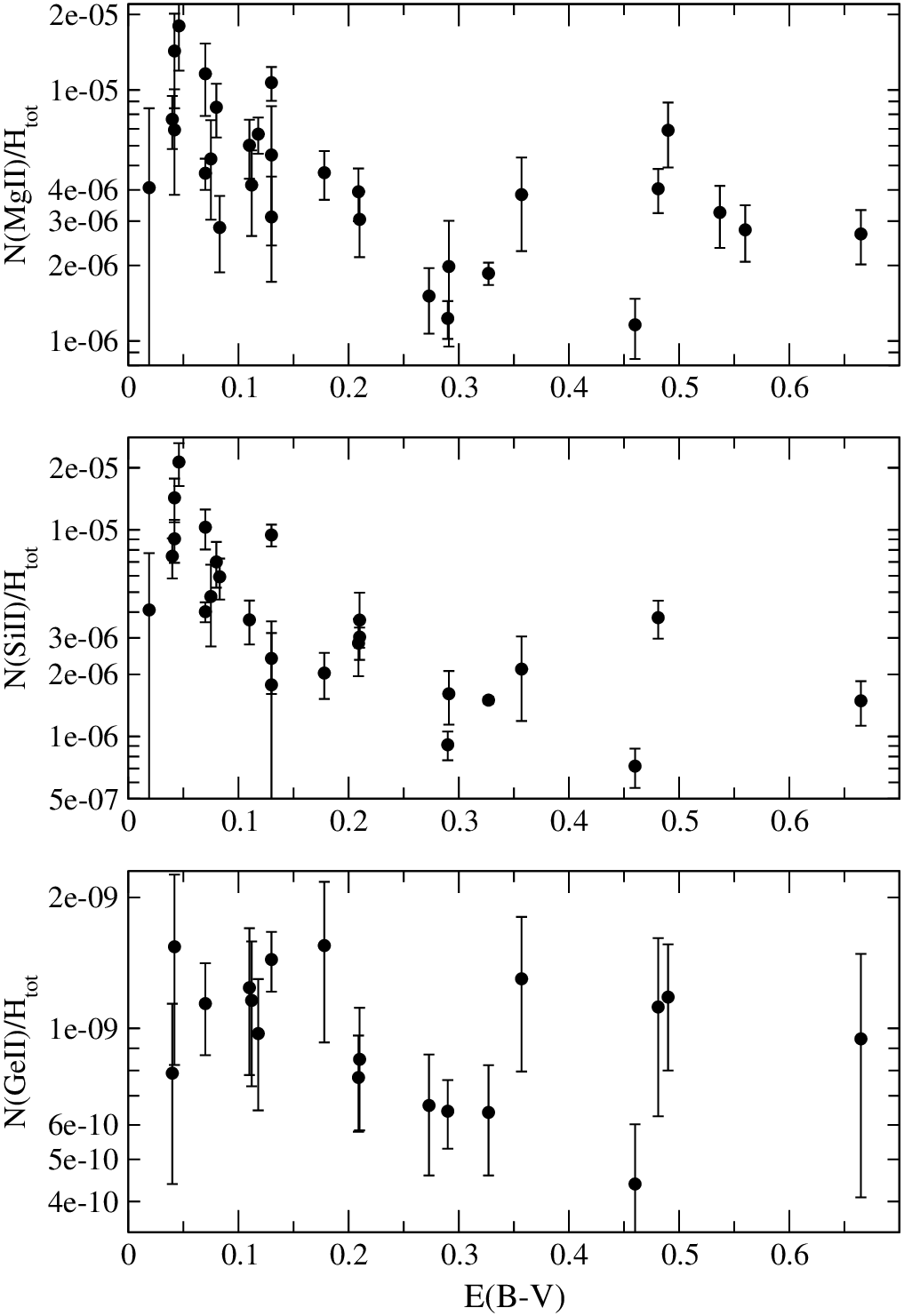}
      \caption{ Column densities of Mg II, Si II and Ge II normalised to the total hydrogen column density (logarithmic scale) versus the interstellar reddening E(B-V).
              }
      \label{NEBV}
\end{figure*}

    The column density of Mg II, Si II and Ge II (dominant ionisation stages) normalised to the total hydrogen column density (see Table \ref{H}) versus interstellar reddening is presented on Fig. \ref{NEBV}. The hydrogen--normalised Si II and Mg II column density drops
for sites with more interstellar dust (grater interstellar reddening E(B-V)). For Ge II the drop is not so evident. This can be attributed to the lower condensation temperature of germanium ($T_c(Ge)=825$ K, versus $T_c(Mg)=T_c(Si)=1340$ K).
    
\begin{figure*}[tbh]
   \centering
   \includegraphics[width=120mm]{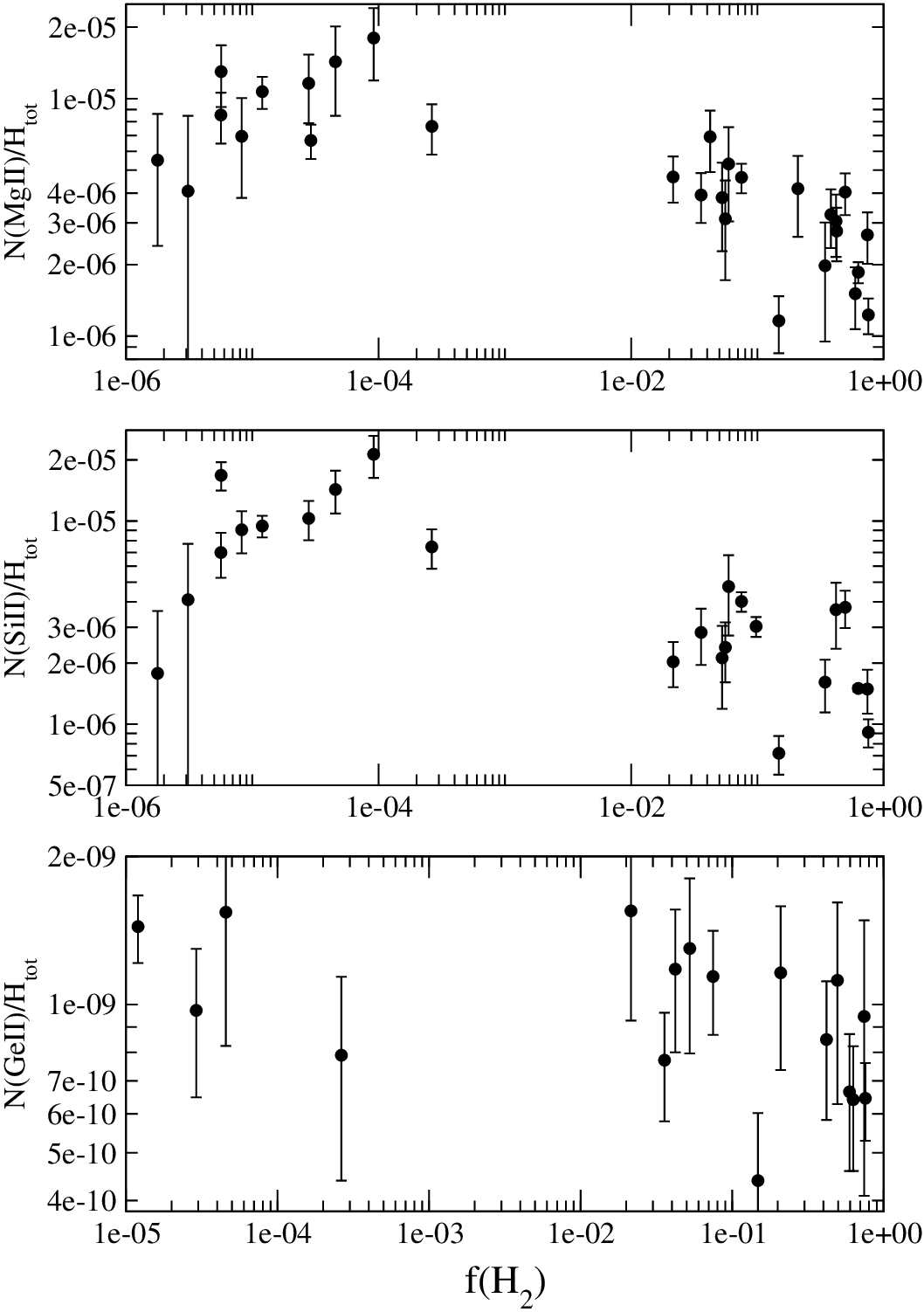}
      \caption{ Column densities of Mg II, Si II and Ge II normalised to the total hydrogen column density versus fractional abundance of molecular hydrogen f(H$_2$).
Both axis use logarithmic scale.
              }
      \label{NfH2}
\end{figure*}

The physical conditions in the interstellar clouds can be characterised by the fractional
abundance of molecular hydrogen $f(H_2)=2N(H_2)/(N(HI)+2N(H_2))$.
The $H_2$ molecule is in equilibrium between formation on dust grains and gas-phase destruction in chemical processes or in photodissociation. 
Some elements , such as the noble gas krypton, do not exhibit changes with varying $f(H_2)$,
because they do not condense onto dust grains (Cardelli, 1994). On the other hand elements that are easily incorporated into dust grains (Mg, Si) show larger depletions in clouds with large $f(H_2)$ (molecular clouds) -- see Fig. \ref{NfH2}.
Only a slight decrease of germanium abundance was noticed by Cardelli (1994) in a limited sample of directions. In our data (Fig. \ref{NfH2}) the increase of germanium depletion is seen in dense, molecular clouds. The errors of Ge abundance are large, due to the weakness of the 1337 \AA\ line.
        
\begin{figure*}[tbh]
   \centering
   \includegraphics[width=120mm]{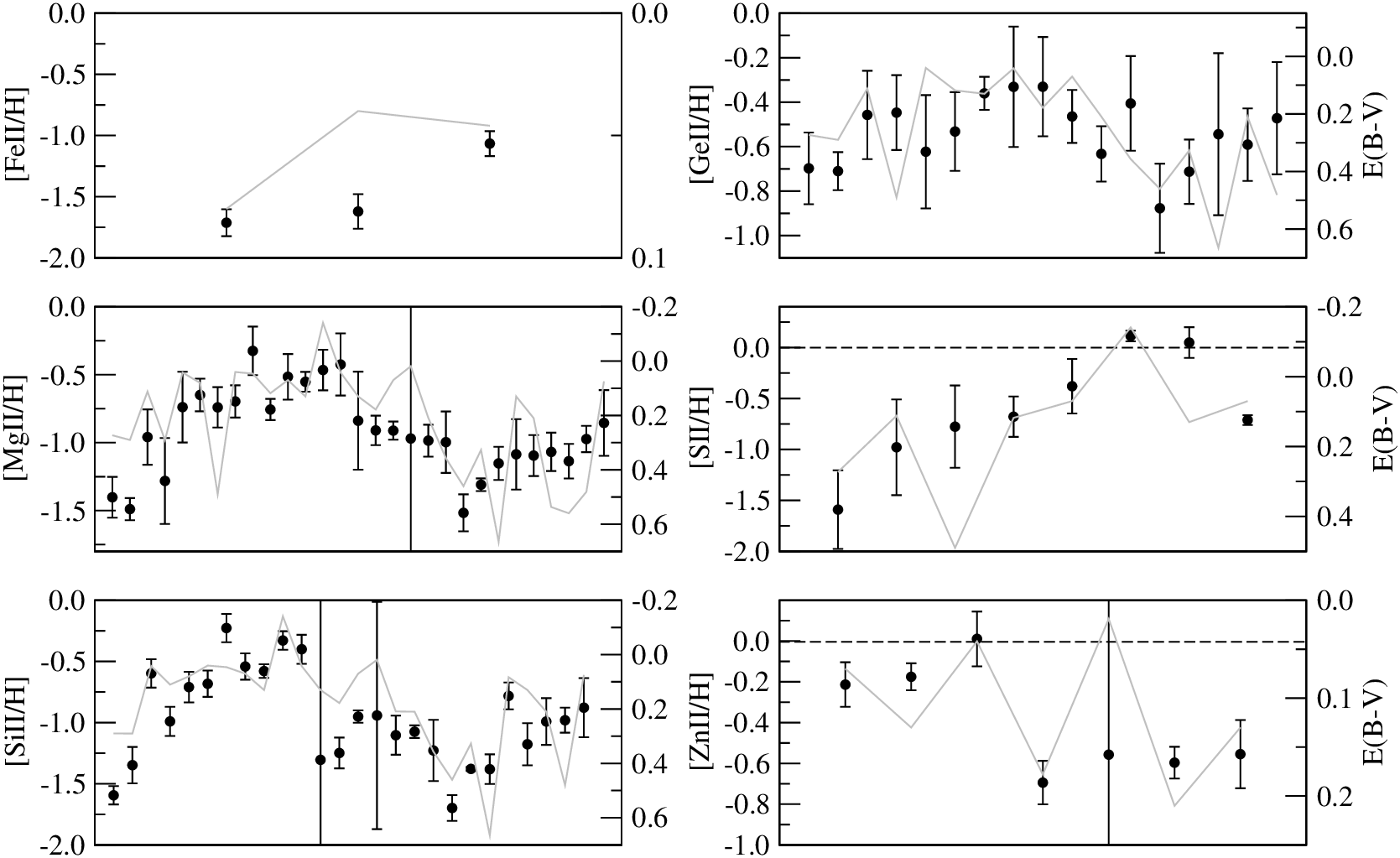}
      \caption{ Black dots show the abundance of the dominant ionization stage relative to the Solar System abundance [dex]. The grey lines represent the interstellar reddening E(B-V). The stars are ordered by HD numbers.
              }
      \label{Obfitosc}
\end{figure*}

    The abundances of elements in the dominant ionisation stage are lower than the Solar System abundances for Fe II, Ge II, Mg II and Si II (Fig. \ref{Obfitosc}). The Solar System abundances were adopted from Grevesse \& Sauval (2000) and are equal to
\begin{math}
(Fe/H)_\odot=3.16\cdot10^{-5} \\
(Ge/H)_\odot=3.31\cdot10^{-9} \\
(Mg/H)_\odot=3.80\cdot10^{-5} \\
(Si/H)_\odot=3.59\cdot10^{-5} \\
(S/H)_\odot=1.85\cdot10^{-5}  \\
(Zn/H)_\odot=4.31\cdot10^{-8} \\
\end{math}
In the direction to HD 91316 the Zn II abundance is grater than the Solar System abundance.
Zinc is known to not deplete into dust grains, although the Zn abundance may be lower than the Solar System abundance ($[Zn/S]=log(ZnII/S)-log(Zn/S)_\odot=-0.13$, Fitzpatrick \& Spitzer (1996)). 
The S II abundance is grater than the Solar System abundance for HD 116658 and HD 68273. 
Sulphur was known to not incorporate into dust grains.
Fitzpatrick \& Spitzer (1996) and Fitzpatrick (1996) have stated that the sulphur 
abundance is close to the solar abundance in diffuse clouds.
In our data only 2 of 8 directions have the sulphur abundance close to the solar abundance. The depletion of sulphur reaches $[S/H]=-1.59$ for HD 24398. 
Such low sulphur abundance leads to doubts if sulphur can be used as a tracer of the total hydrogen column density.
The low depletion of S and Zn is not surprising, since this elements have much lower
condensation temperatures ($T_c(S)=674$ K, $T_c(Zn)=684$ K) than other analysed elements (for Fe, Mg and Si $T_c\sim1340$ K).
The sin--like pattern seen on Fig. \ref{Obfitosc} is caused by changes in the interstellar reddening E(B-V) in the analysed directions.

\begin{landscape}
\tiny
\begin{longtable}{llllllllll}
\hline \hline
Star &  Par & FeII & GeII & MgI & MgII & SI & SII & SiII & ZnII \\
\hline
\endhead
BD+28 4211 & 9.63$\pm$1.71 & --- & --- & $9.18\pm0.91\cdot10^{11}$ & $5.28\pm1.90\cdot10^{13}$ & --- & $6.93\pm0.28\cdot10^{14}$ & $1.28\pm0.05\cdot10^{14}$ & --- \\
Feige 24 & 13.44$\pm$3.62 & --- & --- & --- & --- & --- & $7.53\pm2.64\cdot10^{13}$ & $2.76\pm0.15\cdot10^{13}$ & --- \\
GD246 & --- & --- & --- & --- & --- & --- & --- & $9.80\pm0.20\cdot10^{13}$ & --- \\
HD 432 & 59.89$\pm$0.56 & $2.04\pm0.17\cdot10^{12}$ & --- & --- & $3.65\pm0.04\cdot10^{12}$ & --- & --- & --- & --- \\
HD 11443 & 50.87$\pm$0.82 & $3.31\pm0.77\cdot10^{12}$ & --- & --- & $7.52\pm0.04\cdot10^{12}$ & --- & --- & --- & --- \\
HD 18100 & 0.88$\pm$1.04 & $3.39\pm0.08\cdot10^{14}$ & --- & $1.72\pm0.06\cdot10^{12}$ & --- & --- & --- & --- & --- \\
HD 22049 & 310.75$\pm$0.85 & --- & --- & --- & $3.66\pm0.18\cdot10^{12}$ & --- & --- & --- & --- \\
HD 22468 & 34.52$\pm$0.87 & --- & --- & --- & $2.55\pm0.33\cdot10^{12}$ & --- & --- & --- & --- \\
HD 23630 & 8.87$\pm$0.99 & --- & --- & $1.07\pm0.03\cdot10^{12}$ & $1.06\pm0.15\cdot10^{15}$ & --- & $4.30\pm1.71\cdot10^{14}$ & $8.44\pm0.11\cdot10^{14}$ & --- \\
HD 24398 & 3.32$\pm$0.75 & --- & $1.05\pm0.02\cdot10^{12}$ & --- & $2.38\pm0.003\cdot10^{15}$ & --- & $7.50\pm2.23\cdot10^{14}$ & --- & --- \\
HD 24534 & 1.21$\pm$0.94 & --- & $1.42\pm0.11\cdot10^{12}$ & $7.49\pm0.46\cdot10^{13}$ & $2.71\pm0.20\cdot10^{15}$ & --- & --- & $2.01\pm0.12\cdot10^{15}$ & --- \\
HD 24760 & 6.06$\pm$0.82 & --- & $3.65\pm0.28\cdot10^{11}$ & --- & $1.32\pm0.11\cdot10^{15}$ & --- & $6.14\pm2.27\cdot10^{14}$ & --- & --- \\
HD 24912 & 1.84$\pm$0.70 & --- & --- & $2.76\pm0.14\cdot10^{13}$ & $3.93\pm0.96\cdot10^{15}$ & $1.49\pm0.06\cdot10^{13}$ & --- & $3.19\pm0.06\cdot10^{15}$ & --- \\
HD 28497 & 2.07$\pm$0.82 & --- & --- & $4.86\pm0.11\cdot10^{12}$ & $1.11\pm0.28\cdot10^{15}$ & --- & --- & $1.45\pm0.05\cdot10^{15}$ & --- \\
HD 34029 & 77.29$\pm$0.89 & $2.71\pm0.15\cdot10^{12}$ & --- & --- & $6.77\pm0.14\cdot10^{12}$ & --- & --- & --- & --- \\
HD 35149 & 3.39$\pm$0.87 & --- & $6.83\pm1.16\cdot10^{11}$ & $4.36\pm2.13\cdot10^{13}$ & $3.31\pm0.21\cdot10^{15}$ & $5.78\pm0.68\cdot10^{12}$ & $1.06\pm0.32\cdot10^{15}$ & $2.02\pm0.08\cdot10^{15}$ & $7.43\pm0.09\cdot10^{12}$ \\
HD 36486 & 3.56$\pm$0.83 & $1.04\pm0.02\cdot10^{14}$ & --- & --- & $1.45\pm0.06\cdot10^{15}$ & --- & --- & $1.19\pm0.06\cdot10^{15}$ & --- \\
HD 36861 & 3.09$\pm$0.78 & --- & $7.41\pm0.60\cdot10^{11}$ & --- & $4.33\pm0.21\cdot10^{15}$ & --- & $1.94\pm0.71\cdot10^{15}$ & --- & --- \\
HD 37128 & 2.43$\pm$0.91 & $2.12\pm0.16\cdot10^{14}$ & $2.21\pm0.54\cdot10^{11}$ & $3.46\pm1.68\cdot10^{13}$ & $2.14\pm0.08\cdot10^{15}$ & --- & --- & $2.09\pm0.04\cdot10^{15}$ & --- \\
HD 38666 & 2.52$\pm$0.55 & $1.90\pm0.02\cdot10^{14}$ & --- & $2.91\pm0.47\cdot10^{12}$ & $1.26\pm0.17\cdot10^{15}$ & --- & --- & $1.49\pm0.05\cdot10^{15}$ & --- \\
HD 38771 & 4.52$\pm$0.77 & --- & $3.21\pm0.75\cdot10^{11}$ & --- & $2.20\pm0.14\cdot10^{15}$ & --- & $1.28\pm0.34\cdot10^{15}$ & --- & --- \\
HD 44743 & 6.53$\pm$0.66 & --- & --- & $3.71\pm0.22\cdot10^{11}$ & --- & --- & --- & $2.40\pm0.20\cdot10^{14}$ & --- \\
HD 47839 & 3.19$\pm$0.73 & --- & --- & $7.36\pm0.39\cdot10^{12}$ & $2.90\pm0.35\cdot10^{15}$ & --- & $1.93\pm0.50\cdot10^{15}$ & $2.58\pm0.05\cdot10^{15}$ & $6.59\pm0.15\cdot10^{12}$ \\
HD 57061 & 1.02$\pm$0.71 & --- & $7.22\pm0.41\cdot10^{11}$ & $1.53\pm0.11\cdot10^{13}$ & $5.33\pm0.29\cdot10^{15}$ & --- & --- & $4.73\pm0.10\cdot10^{15}$ & $1.44\pm0.06\cdot10^{13}$ \\
HD 61421 & 285.93$\pm$0.88 & --- & --- & --- & $3.65\pm0.33\cdot10^{12}$ & --- & --- & --- & --- \\
HD 62044 & 26.68$\pm$0.79 & --- & --- & --- & $4.78\pm0.20\cdot10^{12}$ & --- & --- & --- & --- \\
HD 62509 & 96.74$\pm$0.87 & --- & --- & --- & $5.33\pm0.65\cdot10^{12}$ & --- & --- & --- & --- \\
HD 68273 & 3.88$\pm$0.53 & $1.06\pm0.01\cdot10^{14}$ & --- & --- & $7.81\pm1.49\cdot10^{14}$ & --- & $1.44\pm0.02\cdot10^{15}$ & $1.01\pm0.06\cdot10^{15}$ & --- \\
HD 74455 & 1.93$\pm$0.57 & --- & $5.45\pm0.51\cdot10^{11}$ & $1.08\pm0.03\cdot10^{13}$ & $3.07\pm0.01\cdot10^{15}$ & $2.36\pm0.49\cdot10^{12}$ & $3.08\pm0.02\cdot10^{15}$ & $3.51\pm0.06\cdot10^{15}$ & --- \\
HD 89688 & 1.86$\pm$0.80 & --- & --- & --- & --- & $2.35\pm0.25\cdot10^{13}$ & --- & $3.25\pm0.44\cdot10^{15}$ & --- \\
HD 91316 & 0.57$\pm$0.82 & --- & $2.78\pm0.73\cdot10^{11}$ & $5.22\pm1.50\cdot10^{12}$ & $2.57\pm0.54\cdot10^{15}$ & --- & --- & $2.57\pm0.10\cdot10^{15}$ & $7.95\pm0.52\cdot10^{12}$ \\
HD 100340 & 0.70$\pm$1.45 & --- & --- & --- & --- & --- & $1.94\pm0.10\cdot10^{15}$ & --- & --- \\
HD 111812 & 10.62$\pm$0.90 & --- & --- & --- & $3.61\pm0.03\cdot10^{12}$ & --- & --- & --- & --- \\
HD 116658 & 12.44$\pm$0.86 & --- & --- & --- & $5.51\pm1.73\cdot10^{13}$ & --- & $2.07\pm0.09\cdot10^{14}$ & $1.78\pm1.39\cdot10^{13}$ & --- \\
HD 119608 & 0.77$\pm$0.82 & --- & --- & $1.27\pm0.24\cdot10^{13}$ & $7.97\pm1.03\cdot10^{15}$ & --- & --- & $4.60\pm0.27\cdot10^{15}$ & $1.88\pm0.15\cdot10^{13}$ \\
HD 122879 & -0.05$\pm$0.76 & --- & --- & $1.80\pm0.26\cdot10^{13}$ & $1.60\pm0.04\cdot10^{16}$ & --- & --- & $1.22\pm0.05\cdot10^{16}$ & --- \\
HD 141637 & 6.25$\pm$0.91 & --- & $2.45\pm0.50\cdot10^{12}$ & $1.08\pm0.02\cdot10^{13}$ & $7.41\pm0.19\cdot10^{15}$ & --- & --- & $3.21\pm0.18\cdot10^{15}$ & $1.38\pm0.03\cdot10^{13}$ \\
HD 143018 & 7.10$\pm$0.84 & --- & $6.39\pm0.94\cdot10^{11}$ & $1.39\pm0.06\cdot10^{12}$ & $2.62\pm0.13\cdot10^{15}$ & --- & $2.02\pm0.03\cdot10^{15}$ & $2.26\pm0.04\cdot10^{15}$ & --- \\
HD 143118 & 6.61$\pm$0.78 & --- & --- & --- & $4.47\pm1.45\cdot10^{14}$ & --- & --- & $4.50\pm0.60\cdot10^{14}$ & $1.31\pm0.12\cdot10^{12}$ \\
HD 143275 & 8.12$\pm$0.88 & --- & $1.12\pm0.06\cdot10^{12}$ & $1.65\pm0.02\cdot10^{13}$ & $5.71\pm0.25\cdot10^{15}$ & $2.57\pm0.14\cdot10^{12}$ & --- & $4.11\pm0.47\cdot10^{15}$ & --- \\
HD 144217 & 6.15$\pm$1.12 & --- & --- & $1.05\pm0.65\cdot10^{13}$ & --- & $3.20\pm0.47\cdot10^{12}$ & --- & $4.16\pm0.09\cdot10^{15}$ & $1.50\pm0.11\cdot10^{13}$ \\
HD 147165 & 4.44$\pm$0.81 & --- & $3.02\pm0.03\cdot10^{12}$ & $1.58\pm0.17\cdot10^{13}$ & $8.89\pm0.24\cdot10^{15}$ & $3.95\pm0.46\cdot10^{12}$ & --- & $4.93\pm0.29\cdot10^{15}$ & --- \\
HD 147933 & 8.27$\pm$1.18 & --- & $2.20\pm0.46\cdot10^{12}$ & $3.12\pm0.26\cdot10^{13}$ & $5.79\pm0.64\cdot10^{15}$ & $3.52\pm0.63\cdot10^{13}$ & --- & $3.60\pm0.20\cdot10^{15}$ & --- \\
HD 148479 & 5.40$\pm$1.68 & --- & $1.92\pm0.26\cdot10^{12}$ & $4.32\pm0.63\cdot10^{12}$ & $1.67\pm0.08\cdot10^{16}$ & $2.89\pm1.46\cdot10^{12}$ & --- & $6.37\pm0.06\cdot10^{15}$ & --- \\
HD 149499 & 26.94$\pm$1.88 & --- & --- & --- & --- & --- & $1.13\pm0.17\cdot10^{14}$ & $7.39\pm0.44\cdot10^{12}$ & --- \\
HD 149757 & 7.12$\pm$0.71 & --- & $8.98\pm2.38\cdot10^{11}$ & $2.38\pm2.11\cdot10^{13}$ & $2.61\pm0.22\cdot10^{15}$ & $3.42\pm0.46\cdot10^{13}$ & --- & $2.10\pm0.04\cdot10^{15}$ & --- \\
HD 154368 & 2.73$\pm$0.96 & --- & $3.68\pm1.58\cdot10^{12}$ & --- & $1.04\pm0.11\cdot10^{16}$ & $3.41\pm0.92\cdot10^{13}$ & --- & $5.81\pm0.61\cdot10^{15}$ & --- \\
HD 158926 & 4.64$\pm$0.90 & --- & --- & $1.05\pm0.29\cdot10^{11}$ & --- & --- & --- & $1.56\pm0.41\cdot10^{14}$ & --- \\
HD 160578 & 7.03$\pm$0.73 & --- & --- & $8.96\pm2.33\cdot10^{11}$ & $4.39\pm0.57\cdot10^{14}$ & --- & --- & $9.19\pm0.16\cdot10^{14}$ & $4.21\pm0.26\cdot10^{12}$ \\
HD 195965 & 1.91$\pm$0.58 & --- & $1.39\pm0.28\cdot10^{12}$ & $4.67\pm0.84\cdot10^{13}$ & $7.96\pm0.80\cdot10^{15}$ & $1.73\pm0.17\cdot10^{13}$ & --- & $5.56\pm0.14\cdot10^{15}$ & --- \\
HD 198478 & 1.45$\pm$0.55 & --- & $3.19\pm0.25\cdot10^{12}$ & --- & $8.17\pm0.80\cdot10^{15}$ & --- & --- & --- & --- \\
HD 201345 & 0.61$\pm$0.77 & --- & --- & --- & $9.14\pm0.80\cdot10^{15}$ & --- & --- & --- & --- \\
HD 202904 & 3.62$\pm$0.56 & --- & --- & $2.90\pm0.33\cdot10^{12}$ & $1.58\pm0.24\cdot10^{15}$ & --- & --- & $1.21\pm0.04\cdot10^{15}$ & $6.09\pm0.15\cdot10^{12}$ \\
HD 203374 & -0.70$\pm$0.70 & --- & $2.76\pm0.31\cdot10^{12}$ & $1.11\pm0.06\cdot10^{14}$ & $1.22\pm0.05\cdot10^{16}$ & $4.45\pm0.11\cdot10^{13}$ & --- & $1.41\pm0.19\cdot10^{16}$ & --- \\
HD 203664 & 2.23$\pm$1.08 & --- & --- & $1.26\pm0.67\cdot10^{13}$ & $2.93\pm0.59\cdot10^{15}$ & --- & --- & $4.37\pm0.23\cdot10^{15}$ & --- \\
HD 206267 & 2.78$\pm$0.79 & --- & $2.92\pm0.21\cdot10^{12}$ & $1.80\pm0.09\cdot10^{14}$ & $1.05\pm0.05\cdot10^{16}$ & --- & --- & $1.26\pm0.15\cdot10^{16}$ & --- \\
HD 207198 & 1.62$\pm$0.48 & --- & --- & --- & $1.15\pm0.05\cdot10^{16}$ & --- & --- & --- & --- \\
HD 207538 & 0.30$\pm$0.62 & --- & --- & --- & $1.06\pm0.06\cdot10^{16}$ & --- & --- & --- & --- \\
HD 209339 & 0.14$\pm$0.57 & --- & $2.31\pm0.17\cdot10^{12}$ & $1.02\pm0.11\cdot10^{14}$ & $1.06\pm0.02\cdot10^{16}$ & $2.56\pm0.13\cdot10^{13}$ & --- & $9.43\pm1.43\cdot10^{15}$ & --- \\
HD 210839 & 1.98$\pm$0.46 & --- & $3.12\pm0.92\cdot10^{12}$ & $1.43\pm0.06\cdot10^{14}$ & $1.13\pm0.06\cdot10^{16}$ & $4.34\pm0.15\cdot10^{13}$ & --- & $1.05\pm0.07\cdot10^{16}$ & --- \\
HD 212571 & 2.96$\pm$0.71 & --- & --- & $1.29\pm0.77\cdot10^{13}$ & $2.50\pm0.30\cdot10^{15}$ & $2.88\pm0.69\cdot10^{12}$ & --- & $2.23\pm0.04\cdot10^{15}$ & --- \\
HD 219188 & 0.94$\pm$1.16 & --- & --- & $1.36\pm0.18\cdot10^{13}$ & $3.95\pm0.20\cdot10^{15}$ & --- & --- & $3.54\pm0.18\cdot10^{15}$ & --- \\
HZ 43 & 31.26$\pm$8.33 & --- & --- & --- & $1.84\pm0.26\cdot10^{12}$ & --- & --- & --- & --- \\
\hline
\caption{ Parallaxes [mas] and column densities [cm$^{-2}$] for the analysed directions. }
\label{CD}
\end{longtable}
\normalsize
\end{landscape}

\begin{table}
\tiny
\begin{tabular}{llrrrcrrcrrr}
\hline \hline
HD & SpType & E(B-V) & N(HI) & err N(HI) & ref. & N(H$_2$) & err N(H$_2$) & ref. & H$_{total}$ & err H$_{total}$ & f(H2) \\
\hline
BD +28 4211 & Op & - & - & - & - & - & - & - & - & - & - \\
Feige 24 & DAwe… & - & - & - & - & - & - & - & - & - & - \\
GD 246 & DAw… & - & - & - & - & - & - & - & - & - & - \\
HD 432 & F2IV & - & - & - & - & - & - & - & - & - & - \\
HD 11443 & F6IV & - & - & - & - & - & - & - & - & - & - \\
HD 18100 & B5II/III & -0.045 & 1.38E+20 & 3.80E+19 & 2 & - & - & - & - & - & - \\
HD 22049 & K2V & -0.030 & - & - & - & - & - & - & - & - & - \\
HD 22468 & G9V & 0.112 & - & - & - & - & - & - & - & - & - \\
HD 23630 & B7III & 0.063 & - & - & - & 3.46E+19 & 1.18E+19 & 3 & - & - & - \\
HD 24398 & B1Iab: & 0.273 & 6.40E+20 & 6.00E+19 & 4 & 4.70E+20 & 2.00E+20 & 4 & 1.58E+21 & 4.60E+20 & 5.95E-01 \\
HD 24534 & O9.5pe & 0.290 & 5.37E+20 & 6.93E+19 & 5 & 8.32E+20 & 7.32E+19 & 5 & 2.20E+21 & 2.16E+20 & 7.56E-01 \\
HD 24760 & B0.5V+… & 0.112 & 2.50E+20 & 5.00E+19 & 4 & 3.30E+19 & 2.10E+19 & 4 & 3.16E+20 & 9.20E+19 & 2.09E-01 \\
HD 24912 & O7.5IIIe & 0.291 & 1.30E+21 & 3.00E+20 & 4 & 3.40E+20 & 1.20E+20 & 4 & 1.98E+21 & 5.40E+20 & 3.43E-01 \\
HD 28497 & B2V:ne & 0.042 & 1.60E+20 & 3.20E+19 & 1 & 6.60E+14 & - & 1 & 1.60E+20 & 3.20E+19 & 8.25E-06 \\
HD 34029 & G5IIIe+… & -0.060 & - & - & - & - & - & - & - & - & - \\
HD 35149 & B1V & 0.110 & 5.50E+20 & 1.10E+20 & 1 & $<$3.40E+18 & - & 1 & 5.50E+20 & 1.10E+20 & - \\
HD 36486 & O9.5II+… & 0.080 & 1.70E+20 & 3.40E+19 & 1 & 4.80E+14 & - & 1 & 1.70E+20 & 3.40E+19 & 5.65E-06 \\
HD 36861 & O8III & 0.490 & 6.00E+20 & 1.50E+20 & 1 & 1.32E+19 & - & 1 & 6.26E+20 & 1.50E+20 & 4.21E-02 \\
HD 37128 & B0Iab: & 0.040 & 2.80E+20 & 5.60E+19 & 1 & 3.70E+16 & - & 1 & 2.80E+20 & 5.60E+19 & 2.64E-04 \\
HD 38666 & O9.5V & 0.046 & 7.00E+19 & 1.40E+19 & 1 & 3.20E+15 & - & 1 & 7.00E+19 & 1.40E+19 & 9.14E-05 \\
HD 38771 & B0Iab: & 0.118 & 3.30E+20 & 3.30E+19 & 1 & 4.80E+15 & - & 1 & 3.30E+20 & 3.30E+19 & 2.91E-05 \\
HD 44743 & B1II/III & 0.030 & $<$5.00E+18 & - & 1 & $<$2.00E+17 & - & 1 & - & - & - \\
HD 47839 & O7Ve & 0.070 & 2.50E+20 & 5.00E+19 & 1 & 3.50E+15 & - & 1 & 2.50E+20 & 5.00E+19 & 2.80E-05 \\
HD 57061 & O9Ib & 0.130 & 5.00E+20 & 5.00E+19 & 1 & 3.00E+15 & - & 1 & 5.00E+20 & 5.00E+19 & 1.20E-05 \\
HD 61421 & F5IV-V & -0.031 & - & - & - & - & - & - & - & - & - \\
HD 62044 & K1III & 0.050 & - & - & - & - & - & - & - & - & - \\
HD 62509 & K0IIIb & 0.000 & - & - & - & - & - & - & - & - & - \\
HD 68273 & WCvar+… & - & 6.00E+19 & 6.00E+18 & 1 & 1.70E+14 & - & 1 & 6.00E+19 & 6.00E+18 & 5.67E-06 \\
HD 74455 & B1.5Vn & 0.081 & 5.37E+20 & 7.99E+19 & 2 & - & - & - & - & - & - \\
HD 89688 & B3.2IV & 0.107 & - & - & - & - & - & - & - & - & - \\
HD 91316 & B1Iab & 0.042 & 1.80E+20 & 3.60E+19 & 1 & 4.10E+15 & - & 1 & 1.80E+20 & 3.60E+19 & 4.56E-05 \\
HD 100340 & B9 & - & 2.95E+20 & 3.81E+19 & 2 & - & - & - & - & - & - \\
HD 111812 & G0IIIp & -0.023 & - & - & - & - & - & - & - & - & - \\
HD 116658 & B1III-IV+… & 0.130 & 1.00E+19 & 2.50E+18 & 1 & 8.90E+12 & - & 1 & 1.00E+19 & 2.50E+18 & 1.78E-06 \\
HD 119608 & B1Ib & 0.120 & 7.76E+20 & 2.14E+20 & 2 & - & - & - & - & - & - \\
HD 122879 & B0Ia & 0.298 & - & - & - & - & - & - & - & - & - \\
HD 141637 & B1.5Vn & 0.178 & 1.55E+21 & 3.10E+20 & 1 & 1.70E+19 & - & 1 & 1.58E+21 & 3.10E+20 & 2.15E-02 \\
HD 143018 & B1V+… & 0.070 & 5.20E+20 & 5.20E+19 & 1 & 2.10E+19 & - & 1 & 5.62E+20 & 5.20E+19 & 7.47E-02 \\
HD 143118 & B2.5IV & 0.019 & - & - & - & - & - & - & - & - & - \\
HD 143275 & B0.2IVe & 0.209 & 1.40E+21 & 2.80E+20 & 1 & 2.60E+19 & - & 1 & 1.45E+21 & 2.80E+20 & 3.58E-02 \\
HD 144217 & B0.5V & 0.210 & 1.24E+21 & 1.24E+20 & 1 & 6.70E+19 & - & 1 & 1.37E+21 & 1.24E+20 & 9.75E-02 \\
HD 147165 & B1III & 0.357 & 2.20E+21 & 8.80E+20 & 1 & 6.10E+19 & - & 1 & 2.32E+21 & 8.80E+20 & 5.25E-02 \\
HD 147933 & B2/B3V & 0.460 & 4.27E+21 & 7.98E+20 & 2 & 3.70E+20 & - & 1 & 5.01E+21 & 7.98E+20 & 1.48E-01 \\
HD 148479 & B2V & 0.240 & - & - & - & - & - & - & - & - & - \\
HD 149499 & K0V & -0.110 & - & - & - & - & - & - & - & - & - \\
HD 149757 & O9V & 0.327 & 5.20E+20 & 2.60E+19 & 4 & 4.40E+20 & - & 4 & 1.40E+21 & 2.60E+19 & 6.29E-01 \\
HD 154368 & O9Ia & 0.665 & 1.00E+21 & 1.09E+20 & 5 & 1.45E+21 & 2.15E+20 & 5 & 3.89E+21 & 5.39E+20 & 7.43E-01 \\
HD 158926 & B2IV+... & 0.100 & $<$2.40E+19 & - & 1 & 5.00E+12 & - & 1 & - & - & - \\
HD 160578 & B1.5III & 0.083 & 1.55E+20 & 3.19E+19 & 2 & $<$1.70E+14 & - & 3 & 1.55E+20 & 3.19E+19 & - \\
HD 195965 & B0V & 0.225 & 7.94E+20 & 1.49E+20 & 2 & - & - & - & - & - & - \\
HD 198478 & B3Iae & 0.439 & - & - & - & - & - & - & - & - & - \\
HD 201345 & O9p & -0.140 & 7.41E+20 & 1.52E+20 & 2 & - & - & - & - & - & - \\
HD 202904 & B2Vne & 0.130 & 4.79E+20 & 1.40E+20 & 3 & 1.41E+19 & 4.79E+18 & 3 & 5.07E+20 & 1.49E+20 & 5.57E-02 \\
HD 203374 & B0IVpe & 0.513 & 1.29E+21 & 2.41E+20 & 2 & - & - & - & - & - & - \\
HD 203664 & B0.5IIIn & 0.075 & 3.47E+20 & 7.76E+19 & 2 & - & - & - & - & - & - \\
HD 206267 & O6e & 0.210 & 2.00E+21 & 7.00E+20 & 4 & 7.20E+20 & 6.50E+19 & 4 & 3.44E+21 & 8.30E+20 & 4.19E-01 \\
HD 207198 & O9IIe & 0.537 & 2.19E+21 & 7.09E+20 & 5 & 6.76E+20 & 5.95E+19 & 1 & 3.54E+21 & 8.28E+20 & 3.82E-01 \\
HD 207538 & O9V & 0.560 & 2.20E+21 & 6.00E+20 & 4 & 8.10E+20 & 7.60E+19 & 4 & 3.82E+21 & 7.52E+20 & 4.24E-01 \\
HD 209339 & B0IV & 0.304 & - & - & - & - & - & - & - & - & - \\
HD 210839 & O6Iab:… & 0.481 & 1.41E+21 & 2.91E+20 & 5 & 6.92E+20 & 6.09E+19 & 5 & 2.80E+21 & 4.12E+20 & 4.95E-01 \\
HD 212571 & B1Ve & 0.092 & - & - & - & - & - & - & - & - & - \\
HD 219188 & B0.5III & 0.075 & 7.00E+20 & 2.80E+20 & 1 & 2.20E+19 & - & 1 & 7.44E+20 & 2.80E+20 & 5.91E-02 \\
HZ 43 & DAw… & - & - & - & - & - & - & - & - & - & - \\
\hline
\end{tabular}
\caption{ Interstellar reddening and hydrogen column density for target stars. References:
1 - Bohlin {\it et al.} (1978) ; 2 - Diplas \& Savage (1994) ; 3 - Jenkins {\it et al.} (1986) ; 4 - Lacour {\it et al.} (2005) ; 5 - Rachford {\it et al.} (2002). }
\label{H}
\end{table}

\section{Conclusions}

\begin{itemize}
  \item The column density of ionised elements Mg II, S II, Si II, Zn II correlates with the distance.
  \item Estimation of a distance from the column density can lead to large overestimation of the distance for star lying behind molecular clouds. 
  \item Sulphur is depleted in comparison to the Solar System abundance. Sulphur should not be used as a tracer of the hydrogen column density.
  \item Germanium abundance drops with increasing fractional abundance of molecular hydrogen.
\end{itemize}

\bigskip

{\bf Acknowledgements. }
   The authors would like to thank professor Jacek Kre\l owski for helpful suggestions.
   This publication is based on observations made with the NASA/ESA 
Hubble Space Telescope, obtained from the data archive at the Space 
Telescope Science Institute. STScI is operated by the Association of 
Universities for Research in Astronomy, Inc. under NASA contract NAS 5-26555.
This work was supported by University of Gda\'nsk grant BW/5400-5-0219-6.

\begin{center}

\end{center}

\end{document}